\documentclass[11pt]{article}
\usepackage{amsmath}
\usepackage{amssymb}
\usepackage{graphics}

\newcommand{\tr}{\operatorname{tr}}

\newcommand{\Dbar}{\overline{\text{D8}}}

\begin{document}

\title{Sakai-Sugimoto Brane System at High Density}
\author{D.~Yamada
        \bigskip
        \\
       {\it Racah Institute of Physics},
       {\it The Hebrew University of Jerusalem},
        \smallskip
       \\
         {\it Givat Ram, Jerusalem, 91904 Israel}
          \bigskip
       \\
         {\tt daisuke@phys.huji.ac.il}}
\date{}
\maketitle

\begin{abstract}
The D4-D8 brane system of
Sakai-Sugimoto model at high quark density is studied in
the weak coupling regime.
We show that the color superconducting phase
(for $N_c\approx 3$)
or the chiral density wave (for $N_c\to\infty$) disappears 
at very large chemical potential, or equivalently at very
large compactified dimension that the model possesses.
We also comment on the prospects in the strong coupling regime
along with the QCD phase diagram.
\end{abstract}

\bigskip

\tableofcontents
\pagebreak

\section{Introduction}
Decades have passed since quantum chromodynamics was accepted as
the theory of strong interaction.
While the successes of this theory are impressive, there still remain
numerous unresolved problems.
The most prominent of those problems is the confinement.
At low energy, we observe hadrons instead of the QCD fundamental degrees of
freedom and we still do not quite understand the mechanism of this
phenomenon.
The lack of the understanding is mainly due to the strongly coupled
nature of the theory at low energy and to the fact that 
we do not have very good analytical
control over the field theory in such coupling regime.
The best hope, therefore, is that the numerical study could
provide important insights into QCD at strong coupling.

Confinement is one of the phases of QCD and there are other phases
in different regions of the QCD parameter space.
One such example is the deconfined phase at high temperature.
The numerical method, in fact, is proven to be powerful in the investigation
of the confinement/deconfinement phase transition, and has provided the
estimates of the transition temperature
and the order of the phase transition.
(See Reference~\cite{Rajagopal:2000wf} and the papers cited therein.)
There is, however, a large region of the parameter space in which
even the numerical study has not been very successful. 
Namely, the theory at finite quark (baryon) density.
There is a problem in carrying out lattice simulations with the nonzero
chemical potentials that are conjugate to the density and
this is known as the sign problem (see {\it e.g.} 
Reference~\cite{Stephanov:2007fk}).
While the effort to overcome the sign problem vigorously continues,
there have been some developments in analytic approaches.
These include the analysis at asymptotically large chemical
potential and also the use of NJL type models.
The former takes the advantage that the QCD coupling is expected to be weak
at very high density and perturbative computations from the QCD Lagrangian
itself are possible.
The latter models are pure fermionic and quartic
couplings, which reproduce properties of QCD interactions
in a certain degree, are introduced.%
\footnote{
  A good modern review on perturbative high density QCD is
  Reference~\cite{Rajagopal:2000wf}.
  The NJL-model was introduced by Nambu and Jona-Lasinio
  \cite{Nambu:1961tp,Nambu:1961fr} as the model that exhibits the chiral
  symmetry breaking and light hadronic spectrum.
  A good review on the NJL model is Reference~\cite{Hatsuda:1994pi}.
  The model is applied to QCD at finite density
  and reviewed in References~\cite{Shovkovy:2004me,Buballa:2003qv}.
}

What has been emerging from these analyses is the very rich phase structure
of QCD in the parameter space of the temperature and the chemical potential.
(See, for example, Reference~\cite{Alford:2006wn} for the phase diagram.)
In general, a cold and highly dense quark matter is expected to become
a color superconductor \cite{Bailin:1983bm}.
As mentioned before,
when the quark chemical potential $\mu$ is very large, the coupling $g(\mu)$
is small and the excitations near the Fermi surface of the quarks,
particles and holes,
are nearly free, and this naive ground state at high density is known as
the Fermi liquid.
However, the pairs of the particles at the antipodal points of the Fermi
sphere are all degenerate and it costs no free energy to form such a
pair.
Then, if there is an attractive force between the particles (or holes),
the pairing actually reduces the free energy of the system, leading to the
instability of the naive ground state against the formation of the
pair (Cooper pair).
This is known as the Bardeen-Cooper-Schrieffer (BCS) instability
\cite{Bardeen:1957mv} and in their original work for the electron gas
in a solid, the attractive force was provided by the phonon exchange.
For the case of high density QCD,
there is an attractive force in a color channel and it leads to the similar
BCS instability.

In particular,
if we have three massless flavors in the theory, a very interesting form of
the condensate has been found to form \cite{Alford:1998mk}.%
\footnote{
  This model with $N_f=3$ massless quarks is an approximate QCD where
  the masses of $u,d,s$ are set to zero and those of $c,b,t$ are 
  taken to infinity.
}
This is the (scalar) diquark condensate of the form
\begin{equation}
  \langle q_{L i}^a q_{L j}^b \rangle
  =
  -\langle q_{R i}^a q_{R j}^b \rangle
  =
  \Delta_{CFL} (\delta_i^a\delta^b_j - \delta_j^a\delta_i^b)
  \;,
\end{equation}
where the superscripts
$a,b$ are the color indices, the subscripts
$i,j$ are for the flavors, the subscripts $L,R$ indicates
the chirality of the quarks and $\Delta_{CFL}$ is the size of the
condensate (the gap).
As one can see, the Kronecker deltas relate the flavor and color symmetries
and those are not separately preserved by the condensate.
The residual symmetry is the simultaneous flavor and
(global) color rotations and
for this reason, this phenomenon is called the color-flavor locking (CFL).
The full symmetry breaking pattern of CFL is
\begin{equation}\label{eq:CFLBreaking}
  SU(3)_{\text{color}}\times SU(3)_L\times SU(3)_R\times U(1)_V\times U(1)_{EM} 
  \to SU(3)_{\text{color}+L+R}\times\mathbb{Z}_2\times U(1)_{\tilde Q}
  \;,
\end{equation}
where $SU(3)_{\text{color}+L+R}$ is the global diagonal subgroup of the original
color and flavor symmetries, $\mathbb{Z}_2$ is the subgroup of $U(1)_V$
that changes the sign of all the quarks
and $U(1)_{\tilde Q}$ is known as the ``modified electromagnetism''
whose gauge boson is a linear combination of the original photon and one of 
the gluons \cite{Alford:1998mk,Rajagopal:2000wf}.
We observe that
since $L$- and $R$-flavor symmetries both lock to the color, the chiral
symmetry is broken through the color factor.
Even though the mechanism of the chiral symmetry breaking is very unusual,
the corresponding chiral Lagrangian can be built \cite{Casalbuoni:1999wu}.
This novel phase of QCD created a renewed and wide interest in the QCD
phase structure and many generalizations and modifications have been explored.
For example,
when one considers finite quark masses, there are other possible forms of
the condensate and those can be energetically favored for some regions of the
$\mu$-T parameter space, 
resulting in the complicated structure of the phase diagram.
The interested reader can pursuit the subject in the review papers
cited above.

Rather than continuing to overview the QCD phase structure, we would now like to
turn to the high density behavior in the 't Hooft limit with weak coupling.
(This is not QCD, which has $N_c=3$, but has a potential relation
to the holographic theories.)
In this case, we do not expect the color superconductor
to be the correct ground state of the cold QCD.
Heuristically, this is because the Cooper pair is not a color singlet and
not expected to survive the limit.
The possibility of a color singlet condensate of particle and hole
(not anti-particle)
has been investigated by Deryagin, Grigoriev and Rubakov (DGR) in
Reference~\cite{Deryagin:1992rw}.
When the particle and hole at the antipodal points of the Fermi sphere
form a pair, the condensate is not homogeneous nor isotropic but is
a standing wave in a certain direction;
\begin{equation}
  \langle \bar q_L(x) q_R(y) \rangle
  = e^{i\vec p_F \cdot (\vec x + \vec y)} f(x-y)
  \;,
\end{equation}
where $|\vec p_F|=\mu$ and $f$ is a function that describes the amplitude
of the standing wave.
The condensate is called the chiral density wave ($\chi$DW) and it breaks
the chiral symmetry but not the gauge symmetry (in the limit $x\to y$).
This condensate is kinematically less favored
than the Cooper pairing at $N_c\approx 3$,
but it has been shown that the ground state instability due to the formation
of $\chi$DW (DGR instability) dominates over the BCS-type instability
in the large $N_c$ limit \cite{Deryagin:1992rw,Shuster:1999tn}.

The aim of this paper is to examine the high density behavior of
the Sakai-Sugimoto model \cite{Sakai:2004cn,Sakai:2005yt}.
As we will describe in Section~\ref{sec:weakCoupling}, this is a model
in Type IIA string theory with a certain brane configuration.
What makes this model interesting is that the low energy spectrum
is similar to that of QCD, especially, there are
fundamental quarks with the left- and right-chiral symmetries.
In the strong coupling gravity background analysis, Sakai and Sugimoto
have shown that the chiral symmetry is broken and they were able to
compute the hadron spectrum of the theory.
Moreover, they have constructed the low energy effective action of
the theory with the Skyrme term whose soliton excitations can be considered as
baryons.
Those promising successes make this model an interesting candidate
for holographic QCD.
Therefore, we are naturally motivated to examine the model at high
density and ultimately, we hope the model to give insight into the
structure of the QCD phase diagram in all values of the parameters,
especially in the medium density region where the theory is strongly
coupled.

Though we do not get this far in this work, we show in 
Section~\ref{sec:weakCoupling} that the
weak coupling regime of the theory at high density and zero temperature
behaves very similar to QCD with a few differences.
One of the differences is that when the chemical potential is very
large with respect to the compactification size of the model, the
BCS and DGR instabilities at $N_c\approx 3$ and $N_c\to\infty$, respectively,
are absent and the ground state is the Fermi liquid.
In Section~\ref{sec:discussions}, we comment on the known finite density
analysis of the model at strong coupling, contrasting to the QCD expectations.
In the strong coupling analysis, the possibilities of the superconductivity 
and $\chi$DW have not been addressed
and we discuss some prospects in this direction.

\section{Weak Coupling Field Theory Analysis}\label{sec:weakCoupling}
In this section, we discuss the Sakai-Sugimoto brane system at high
density with the weak Yang-Mills coupling of the world-volume theory.
Though the results are relatively straightforward, the computations
are somewhat involved.
We therefore split the discussion into the qualitative and quantitative
parts.
In the first part, we qualitatively explain the high density behavior
of the system, then in the second part, we carry out the computations
and confirm the qualitative expectations.

\subsection{Qualitative Discussion}\label{subsec:qualitative}

\subsubsection{BCS and DGR Instabilities}
Let us briefly review the BCS and DGR instabilities of high density QCD
in a way that would provide the conceptual background for the
quantitative calculations.%
\footnote{
  For a modern exposition of the BCS instability, see Polchinski's TASI
  lecture notes \cite{Polchinski:1992ed}.
}
For simplicity, we set all current quark masses to zero, which is a good
approximation when the quark chemical potential is much larger than the
mass of the heaviest quark.
In the presence of the quark chemical potential, $\mu$, at zero temperature,
we have a well-defined Fermi sphere of radius $\mu$.
As a convention, we take the excitations near the Fermi surface be
particles as opposed to anti-particles.
When $\mu$ is very large,
the anti-particles are buried deep in the Dirac sea and will not
play a role in the following discussion.

On the Fermi surface, the free energy of the states are zero (more precisely,
at the minimum) and it
costs no free energy to change momentum
along the Fermi surface. 
Therefore, the energy scales only in the radial direction of the sphere and
we consider the renormalization group flow as we scale the energy
down toward the Fermi surface.
The relevant degrees of freedom are the particles and holes near
the Fermi surface and we are interested in the effective theory that describes
the dynamics of those excitations.
Let us specify a point on the Fermi surface by the momentum $\vec p_F$
with the magnitude $|\vec p_F|=\mu$ and
decompose a four-momentum near this point as
\begin{equation}\label{eqn:decomposition}
  p^\nu = (E,\, \vec p_F + \vec l_{\parallel} + \vec l_\bot)
  \;,
\end{equation}
where $\vec l_{\parallel}$ is parallel to $\vec p_F$ and $\vec l_\bot$ is
perpendicular to it.
As stated before, only $l_{\parallel}$ scales with energy and $l_\bot$ may be
trivially integrated along the Fermi surface in a diagram computation.
We, therefore, have the $1+1$-dimensional effective theory that describes
the dynamics of the particles and holes.%
\footnote{
  A rigorous derivation of the high density effective theory of QCD was
  carried out by Hong \cite{Hong:1999ru}.
}
It is important to notice that the kinematics is restricted because the 
dynamics must take place near the Fermi surface.
When we consider a particle-particle or hole-hole scattering, it is
clear that the scattering must be near back-to-back, that is, the
scattering pairs must be at antipodal points of the Fermi sphere.
In the back-to-back scattering, the scattering angle may be arbitrary
without spoiling the kinematic restriction and hence the phase space
of this scattering is all over the Fermi surface.

Now in a two-dimensional theory, irrelevant operators of four dimensions
may become relevant or marginal.
In particular, a four-fermion interaction is marginal in two dimensions.
If the interaction is attractive,
the quartic coupling grows as we scale
the energy down toward the Fermi surface and it eventually hits the Landau
pole.
This implies that the perturbation theory breaks down at the infrared
scale around the pole.
For the scattering of particle or hole pairs,
this indicates that the naive ground state of the weakly interacting
particles and holes near
the Fermi surface, the Fermi liquid, is unstable against the formation
of Cooper pairs.
This is the BCS instability, and the new ground state has a gap 
due to the formation of the condensate whose
size is roughly the location of the Landau pole.
Technically, as we will see in the next subsection, the instability is closely
related to the infrared divergence that appear in the perturbation
theory and the gap properly provides the infrared cutoff.
In weakly coupled QCD, {\it i.e.}, QCD at very high density,
the leading order contribution to the interaction is given by the
one-gluon exchange and it is attractive in the antisymmetric
$\bar{\mathbf{3}}$-channel, resulting in the color superconductivity.
\\

We now turn to the DGR instability.
This is associated with the scattering of the particle and hole
located at the antipodal points of the Fermi sphere.
In this case, the scattering is not back-to-back, but near forward.
Unlike the back-to-back case, the scattering angle cannot be
too large to stay near the Fermi surface and the phase space of the
forward scattering is limited to a very tiny patch on the
Fermi surface.
The difference in the phase spaces for the BCS and DGR cases will
be important in the quantitative computations.

Now the particle-hole forward scattering
is similar to the Bhabha scattering whose amplitude has the forward
enhancement.
Thus also in our case, we can expect an
infrared divergence (the DGR instability)
as the exchange gluon becomes very soft,
leading to a condensate of the particle-hole pair ($\chi$DW).
This, however, does not happen in weakly coupled QCD.
The reason is that the finite density screening effect completely
overwhelms such a condensate; it is the screening that provides the
infrared cutoff and not the formation of a condensate.

\begin{figure}[h]
{
\centerline{\scalebox{0.8}{\includegraphics{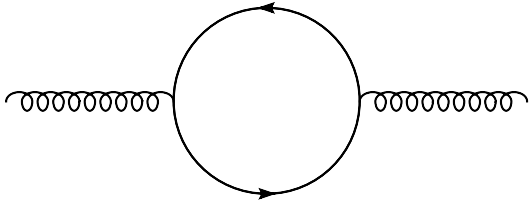}}}
\caption{\footnotesize
  One loop diagram that contributes to the finite density screening.
    }
\label{fig:quarkLoop}
}
\end{figure}

The situation is different in the large $N_c$ limit
(with small 't Hooft coupling).
First, note that the Cooper pair of the BCS instability is not
color singlet while the $\chi$DW of DGR is.
Therefore, the BCS instability is $1/N_c$ suppressed and DGR is not.
Secondly, since only the quarks are charged under the $U(1)_V$-symmetry,
the finite density screening effect is provided by the quark
loops, such as the one shown in Figure~\ref{fig:quarkLoop}, and the
gluon loops do not contribute because the gluon propagator is independent
of the quark chemical potential.
This implies that the finite density screening is suppressed as the
number of the color is taken to a large value.
In fact, the DGR instability was discovered by disregarding the
the screening effect and the authors of Reference~\cite{Deryagin:1992rw}
noted that the $\chi$DW can form and dominate over the Cooper pairing
at least in the large $N_c$ limit.
Later, Shuster and Son showed that the DGR instability may occur
if $N_c\gtrsim 1000N_f$, where $N_f$ is the number of the flavor
\cite{Shuster:1999tn}.

This is an example where the large $N_c$ limit of QCD yields qualitatively
different properties.
In weakly coupled QCD, $\chi$DW is not a relevant phenomenon.
Though it could possibly compete with the Cooper pair at strong coupling,
so far the situation is unclear \cite{Park:1999bz}.

\subsubsection{Sakai-Sugimoto Brane System at High Density}
We now consider whether the Sakai-Sugimoto model at high density exhibits
similar properties as discussed above.
For this purpose, we assume that the model is in the regime with low energy
and weak Yang-Mills coupling so that we can use the perturbative world-volume
field theory arguments.

\begin{figure}[h]
{
\centerline{\scalebox{0.6}{\includegraphics{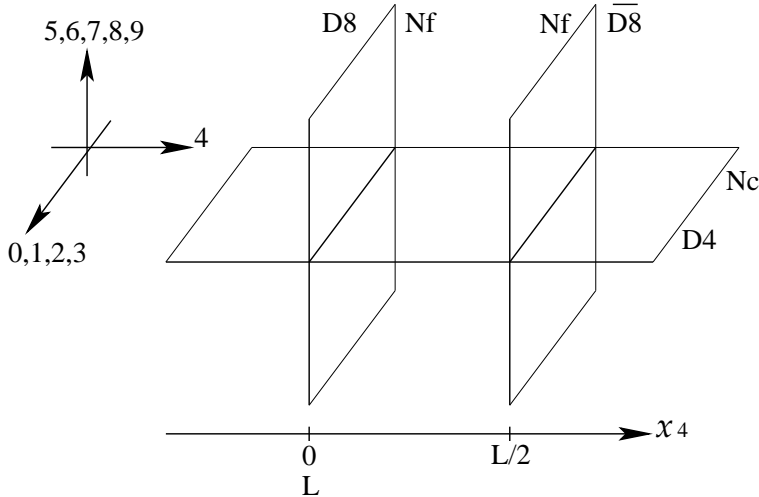}}}
\caption{\footnotesize
  Sakai-Sugimoto brane configuration.
  The $x_4$-direction is compactified with period $L$.
  The D4 world-volume fermions have the anti-periodic boundary condition.
  We locate the $N_f$ D8-branes and  $N_f$ $\Dbar$-branes
  at $x_4=0=L$ and  $x_4=L/2$, respectively.
    }
\label{fig:braneSetup}
}
\end{figure}

The model is Type IIA string theory with D4-, D8- and 
$\Dbar$-branes.
The configuration  of the branes is shown in Figure~{\ref{fig:braneSetup}}.
The $x_4$-direction is compactified to the circle of circumference $L$
and the D8$\Dbar$-branes are placed at the antipodal points of the circle.
To discuss the low energy spectrum of the model, we first consider only
the compactified $N_c$ D4-branes.
In Reference~\cite{Witten:1998zw}, 
Witten suggested that if we impose the anti-periodic
boundary condition to the adjoint world-volume fermions,
the low energy world-volume theory has the spectrum of $3+1$-dimensional
pure Yang-Mills theory.
This is because the fermions get tree level mass of order $1/L$ and
the scalars, including the compactified component of gauge field, $A_4$,
get the one-loop mass of order $g/L$, where $g:=g_5/\sqrt{L}$ and
$g_5^2:=(2\pi)^2g_sl_s$, with $g_s$ string coupling and $l_s$ string
length scale, and we assume $g\ll 1$.%
\footnote{
  Unlike four-dimensional case,
  other components of the gauge field do not acquire mass of order
  $g^2/L$, in all orders of the perturbation theory.
}

Now, Sakai and Sugimoto insert the $N_f$ D8 and $N_f$
$\Dbar$ branes as shown.
As explained in Reference~\cite{Sugimoto:2004mh},
there are massless fermions at the $3+1$-dimensional intersection
of D4- and D8-branes.
These are the lowest states of the strings stretching from D8 to D4.
Since the world-volume $U(N_f)_L$ gauge symmetry of D8-branes acts
as the flavor symmetry, these massless fermions are fundamental
``quarks''.
Similar massless fermions are also present at the intersection of
D4- and $\Dbar$-branes.
However, the GSO projection projects out opposite chiralities to
those fermions at different intersections.
Therefore, we call the massless fermions at D4D8 and 
D4$\Dbar$ intersections as ``left-handed ($q_L$) and 
right-handed ($q_R$) quarks'',
respectively.

This theory at low energy,
therefore, has $U(N_f)_L\times U(N_f)_R$ flavor symmetry
and the theory appears to be very similar to the massless QCD,
if $N_c$ and $N_f$ are appropriately chosen.
The difference, of course, is that the fermions with different
chiralities are separately located in the $x_4$-direction and
the gluons propagate in five dimensions, including the $x_4$-direction.
As argued by Antonyan {\it et al.} \cite{Antonyan:2006vw},
this QCD-like dynamics of the quarks
at the intersections
does not change even if the period $L$ is large;
the shift symmetry of the D4 adjoint scalars and the five dimensional
gauge symmetry allow the scalars, including $A_4$, to couple
to the fundamental fermions only through derivative interactins
which is suppressed by the string scale.

In the weak coupling analysis that we carry out in this section,
we assume that the period $L$ is much larger than the string length
scale $l_s$ so that the tachyon becomes heavy and decouples.%
\footnote{
  When the D4-branes of the system are replaced with their
  effective geometry (which is not the case in our discussion),
  the proper distance of the D8s becomes less than the string
  scale in the region sufficiently near the horizon and
  the statement made here is no longer valid in the analysis
  with the background geometry.
  Refs.\cite{Bergman:2007pm,Dhar:2007bz,Dhar:2008um} 
  show that the tachyon indeed condenses and it
  is responsible for the chiral symmetry breaking of the model
  at strong coupling.
}
Also as we have already explained, we place the D8 and
$\Dbar$ branes at the antipodal points of the compactified circle.
\\

In examining this model, we need to decide on
how the quarks interact through the exchange of the gluons that
propagate in the $x_4$-direction.
One possibility is the non-local interaction.
This scenario takes only into account of the zero-mode of the discrete
momentum in the $x_4$-direction.
In this case, the theory becomes completely insensitive to the
existence of the fifth dimension and behaves in the same way 
as the four dimensional QCD-like theory.
We find this non-local scenario less appealing, especially when the
compactification scale $L$ is large.
We thus take the second alternative where the D8-branes are treated
as sources (or stiff walls) and allow the exchanged gluons to carry
arbitrarily high momenta in the $x_4$-direction.%
\footnote{
  The gluon momenta actually have to be cut off below the string scale
  to avoid the derivative interactions between the D4 adjoint scalars
  and the fundamental fermions.
  However, as we will see shortly, the high momenta increasingly
  suppress the gluon propagator and their contributions become
  negligible at sufficiently high scale.
  Therefore we can approximately take the $x_4$-momentum to infinity.
}
Such an assumption is reasonable because the D8 branes are infinitely
heavier than the D4s and consistent with the fact that we are
treating the D8s as the flavor branes, that is, we are neglecting
their fluctuations.
The momentum is not conserved in the $x_4$-direction but this is
natural in that the translation symmetry is broken in this direction.

We now explain how we introduce the quark chemical potential.
We have the global symmetry $SU(N_f)_L\times SU(N_f)_R\times U(1)_V$.
The chemical potential that we are interested in is conjugate to the
$U(1)_V$ charge.
The standard way to introduce a chemical potential in a field theory
is to treat it
as the constant background ``gauged'' field of a $U(1)$ global symmetry,
with all the components being zero except the time component.
In this way, the chemical potential modifies the time component
of the covariant derivatives in the Lagrangian of the theory.
Thus in our case, the simplest way to introduce the quark chemical potential
is to turn on the $A_0$ constant
background gauge fields of $U(1)\in U(N_f)_L$ and $U(1)\in U(N_f)_R$
world-volume gauge symmetries and tune them to an equal value.
Actually, the background fields may not be constant all over the D8-branes
and the only requirement is to have the constant value at the intersections
with the D4-branes.
Thus, for example, we may turn on the field that depends on the radial
direction in the 5,6,7,8 and 9 directions which would correspond to
the nonzero electric field in the world-volume.
\\

What we will find in the quantitative analysis of the next section is
rather intuitive.
As we have mentioned, if we consider the effect only of the zero-mode
momentum in the $x_4$-direction, the theory reduces to the QCD-like theory.
Thus when the compactification scale $1/L$ is very large compared
to the energy scale of the interest, in this case it is the value of the
chemical potential $\mu$, we expect to have the BCS and
DGR instabilities at $N_c\approx 3$ and at $N_c\to\infty$, respectively,
just as described before.
Now as the compactification scale $1/L$ gets smaller, the infrared effect
in the $x_4$-direction becomes comparable to the one that leads to the
regular BCS or DGR instability.
As a consequence, the size of the condensate grows
and eventually becomes too large to maintain the dynamics near the
Fermi surface.
Therefore, when the scale $1/L$ is small with respect to $\mu$,
there is no BCS or DGR type of instability and the
ground state of the theory is described by the Fermi liquid.

\begin{figure}[h]
{
\centerline{\scalebox{0.5}{\includegraphics{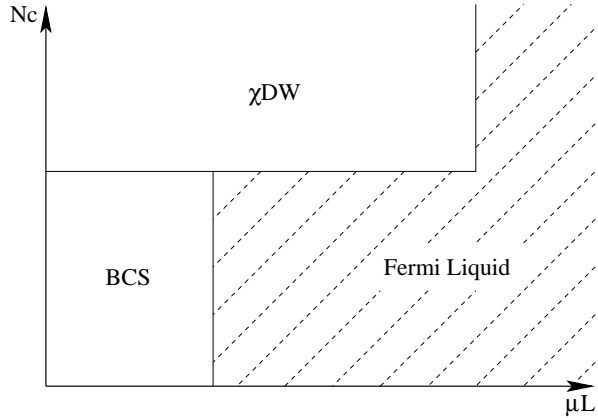}}}
\caption{\footnotesize
  A schematic phase diagram of the theory at weak coupling, high density
  and zero temperature.
  Being schematic, the straight lines may not be straight
  nor sharp transitions in reality.
    }
\label{fig:schematic}
}
\end{figure}

In this qualitative discussion, it is not clear at what scale this
crossover occurs.
The computations of the next subsection show that at $N_c\approx 3$ and
$\mu L\gtrsim 1/g$, 
no BCS-type instability is present and at $N_c\to\infty$ and 
$\mu L\gtrsim e^{1/\sqrt{\lambda}}/\sqrt{\lambda}$ with $\lambda := g^2N_c$,
no DGR-type instability happens.
Notice that the DGR instability persists to exponentially larger value of
$\mu L$ compared to the BCS case.
This is because the phase space of the particle-hole scattering is very
small, in fact it is exponentially small, and the discrete momentum
in the compactified direction must become as fine as this scale to
open up the extra dimension.
The situation explained here is {\it schematically} summarized in
Figure~\ref{fig:schematic}.

\subsection{Quantitative Discussion}\label{subsec:quantitative}
We now demonstrate quantitatively what has been discussed in the
previous subsection.
We carry out the renormalization group and Dyson-Schwinger analyses.
The former is more intuitive in accordance with the qualitative
discussion and shows the existence of the instabilities.
But this method does not provide the size of the gap and this
is augmented by solving the Dyson-Schwinger equations.

We adopt the conventions of Wess and Bagger~\cite{book:WB}, except
the definition of the Dirac spinor;
\begin{equation}\label{eq:Dirac}
  \psi := \binom{q_{L\alpha}}{q_R^{\;\ \dot\alpha}}
  \;,\quad
  \bar\psi := (\bar q_R^{\;\ \alpha}, \bar{q}_{L\dot\alpha})
  \;.
\end{equation}
We mainly work in the chiral basis.
As a convention, the undotted and dotted spinors live on the D8 and $\Dbar$
branes, respectively.

Our central focus of this subsection is to show the existence of the
instabilities and to obtain the size of the gap.
We are less interested in the exact color-flavor
structure of the condensate,
so in what follows, we simplify the analysis by
suppressing the flavor structure.
This is similar to $N_f=2$ case where the Pauli principle requires
the simpler quark pairing.

\subsubsection{Renormalization Group Equations}
Our first analysis is macroscopic in a sense that we introduce
an effective one point four-fermion coupling.
Then we observe how the effective coupling evolves as we scale the
energy of the system down to the Fermi surface.
This idea was first carried out in high density QCD by
Evans {\it et al.} in Reference~\cite{Evans:1998ek}.

\begin{figure}[h]
{
\centerline{\scalebox{0.8}{\includegraphics{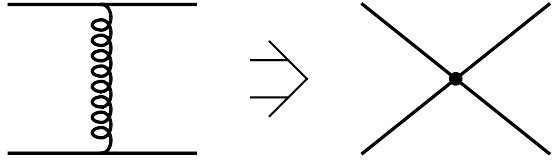}}}
\caption{\footnotesize
  Replacing the one-gluon exchange to an effective one point interaction.
    }
\label{fig:onePoint}
}
\end{figure}

Since we are dealing with the weak coupling at high density,
the quark interaction can be approximated by a single gluon exchange.
We then model the four-fermion interaction by replacing
the one-gluon exchange to a point, as shown in Figure~{\ref{fig:onePoint}}.
Because the chemical potential breaks the $3+1$ world-volume
symmetry down to $O(3)$,
we separately handle the couplings, $G^0$ and $G^j$, as in
\begin{equation}\label{eqn:0jsplit}
  iG^0(\bar\psi\gamma^0\psi)^2\;,\quad iG^j(\bar\psi\gamma^j\psi)^2
  \;.
\end{equation}
For the one-gluon exchange, we can further write
\begin{equation}\label{eqn:defnG}
  G^0(D) = -G^j(D) := -g_5^2 X(D) F
  \;.
\end{equation}
Notice that we have included the minus sign from the signature in the
definition of $G^0$.
The constant $g_5$ is the five dimensional Yang-Mills coupling as before,
$F$ is the form factor that arises from the gluon propagator and
\begin{equation}
  X(D) := \frac{1}{2} \left\{ C(D) - 2C(\Box) \right\}
  \;,
\end{equation}
with $C(D)$ being the Casimir operator of $SU(N_c)$ in the representation
$D$ and $\Box$ being the defining representation.

\begin{figure}[h]
{
\centerline{\scalebox{1.0}{\includegraphics{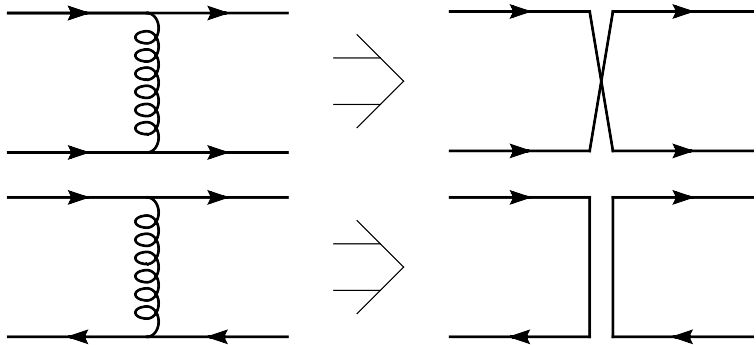}}}
\caption{\footnotesize
  The double line notations of the one-gluon exchange.
  The top and bottom sets represent the symmetric (or antisymmetric)
  and singlet channels, respectively.
  If we fix the colors of the incoming quarks, the symmetric channel
  has the fixed colors for the scattered quarks, while the singlet channel
  has $N_c$ choices.
    }
\label{fig:doubleLine}
}
\end{figure}

We consider three color channels; symmetric (symm), 
antisymmetric (asymm) and singlet ($\bullet$).
For those cases, we have
\begin{equation}\label{eqn:colorFactors}
  X(\text{symm}) = \frac{N_c-1}{2N_c}
  \;,\quad
  X(\text{asymm}) = -\frac{N_c+1}{2N_c}
  \;,\quad
  X(\bullet) = -\frac{N_c^2-1}{2N_c}
  \;,
\end{equation}
where we have adopted the normalization 
$\tr(T^\alpha T^\beta)=(1/2)\delta^{\alpha\beta}$ for
the $N_c\times N_c$ matrices $\{T^\alpha : \alpha =1,\ldots, N_c^2-1\}$ 
of the defining representation.
Notice the large $N_c$ behavior of those factors.
The singlet channel is larger than the other channels by a factor of $N_c$
in the absolute value.
This can be easily understood in the double line notation of the single
gluon exchange diagrams, as shown in Figure~\ref{fig:doubleLine}.

\begin{figure}[h]
{
\centerline{\scalebox{1.0}{\includegraphics{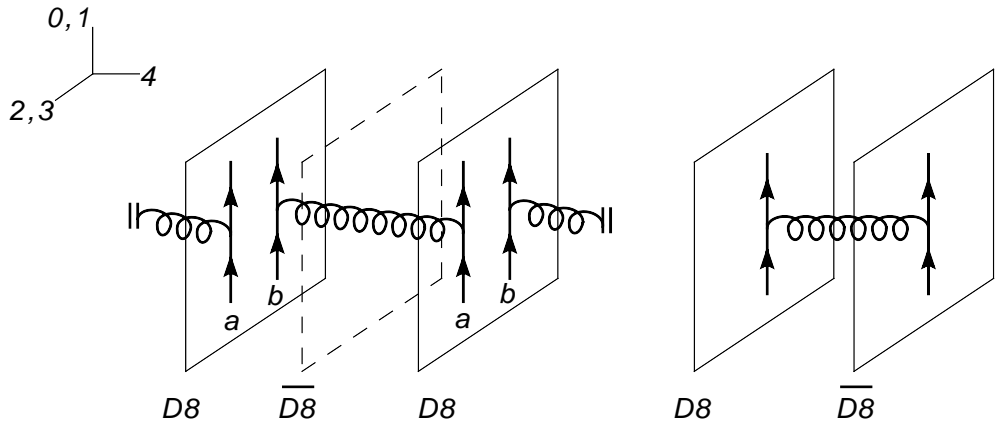}}}
\caption{\footnotesize
  One-gluon exchange diagram.
  The gluons are propagating in the compactified $x_4$-direction.
  The D8$\Dbar$-branes are treated as sources or stiff walls.
    }
\label{fig:interBrane}
}
\end{figure}

Let us now consider the form factor, $F$.
The gluon propagator in the Feynman gauge has the form $1/(p^2+p_4^2)$ with
the discrete momentum, $p_4=2\pi n/L$, in the compactified direction
and this propagator suffers from an infrared divergence.
To cure this problem, we introduce an infrared cutoff $m$.
We simplify the situation by assuming that $m$ is the same for the time
and spatial components of the gluons.%
\footnote{
  At very high density where the coupling is weak, one can expect the
  finite density Debye screening of order $g\mu$
  provides the infrared cutoff.
  This is true for the time component but not for the spatial components.
  There is no static screening in the spatial (magnetic) components
  and the magnetic screening is dynamical due to the Landau damping.
  This was pointed out by Son~\cite{Son:1998uk} and he discovered that
  the dynamical screening effect leads to the qualitatively different
  form of the gap.
  We will take into account of Son's effect in the Dyson-Schwinger
  analysis.
}
Possible one-gluon interactions are illustrated in Figure~\ref{fig:interBrane}.
In both cases shown in the figure, the discrete momentum is $2\pi n/L$
because this is determined by the period of the compactification and not
by the distance between the branes.
Therefore, the form factor, that arises from the gluon propagator,
is independent of the D8 brane distance
and is the same for the both interactions in Figure~\ref{fig:interBrane}
(including the case with the gluons making many rounds in the compactified
circle).
We follow Evans {\it et al.} \cite{Evans:1998ek} to obtain the form
factor, namely, we take average of the gluon propagator over the 
scattering phase space.
For the dynamics very near the Fermi surface, the energy and the component
of the momentum in the radial direction of the Fermi sphere are almost zero.
Hence apart from the extra $x_4$-direction,
the phase space is two dimensional along the Fermi surface.
Let $p_1$ and $p_2$ be the incoming and outgoing momenta of a scattering
quark.
We then have the gluon momentum
\begin{equation}
  p^2 = (p_1-p_2)^2 \approx 2\mu^2(1-\cos\theta)
  \;,
\end{equation}
where $\theta$ is the scattering angle.

For the back-to-back scattering, the angle ranges from $0$ to $\pi$
and the phase space is all over the Fermi surface.
Therefore, the form factor in this case is
\begin{align}\label{eq:FBB}
  F_{BB} &= \frac{1}{\mathcal{N}L} \sum^\infty_{n=-\infty}
        \int \frac{d^2p}{p^2 + (2\pi n/L)^2 + m^2}
  \nonumber\\
        &=\frac{2\pi}{\mathcal{N}L}
          \ln\left[ \sinh\left(\frac{1}{2}mL\sqrt{1+4\mu^2/m^2}\right)/
                        \sinh\left(\frac{1}{2}mL\right) \right]
  \;,
\end{align}
where we have defined the total phase space factor $\mathcal{N}:=4\pi\mu^2$.
We have already carried out the sum over the $p_4$-discrete momentum because
the $p_4$ dependence is only in the gluon propagator.
As an effect, we have encoded all the information about the extra dimension
into the form factor.
For the forward scattering of a particle-hole pair,
the angle should not take very large value so
that the dynamics takes place near the Fermi surface.
Thus we require, in this case, the angle
range from $0$ to $\theta_{UV}$, where the latter angle is much less
than $\pi$ and limits the phase space to a little patch on the Fermi surface.
The form factor then takes the form
\begin{align}\label{eq:FFW}
  F_{FW} &= \frac{1}{\mathcal{M}L} \sum^\infty_{n=-\infty}
        \int \frac{d^2p}{p^2 + (2\pi n/L)^2 + m^2}
        \nonumber\\
        &=\frac{2\pi}{\mathcal{M}L}
          \ln\left[
            \sinh\left(\frac{1}{2}mL\sqrt{1 +
                  2(1-\cos\theta_{UV})\mu^2/m^2}\right)/
            \sinh\left(\frac{1}{2}mL\right) \right]
  \;,
\end{align}
where $\mathcal{M}:=2\pi\mu^2(1-\cos\theta_{UV})$.
Notice that since $\theta_{UV}\ll\pi$, 
we have $\mathcal{M}F_{FW}<\mathcal{N}F_{BB}$.
This fact will lead to the dominance of the BCS-type instability over
the DGR type for $N_c\approx 3$.
When the dimensionless parameter $mL$ is small, the form factors behave
logarithmically with respect to the parameter $\mu/m$ and they behave
linearly when $mL$ is large.
This is because when the the compactified dimension is very small,
the contributions from $n\neq 0$ is also small and the integral is
effectively two dimensional, leading to the log behavior.
When the compactification size is very large, the discrete momentum
becomes finer and the integral essentially becomes three dimensional
and the form factors behave linearly.
\\

\begin{figure}[h]
{
\centerline{\scalebox{1.0}{\includegraphics{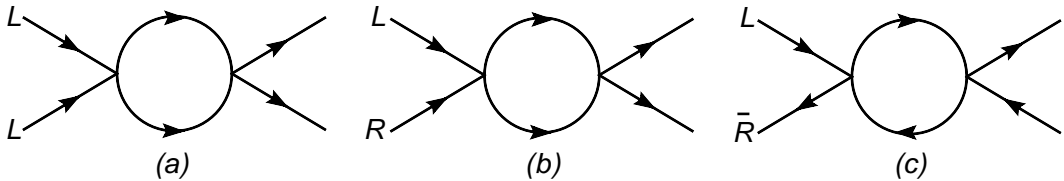}}}
\caption{\footnotesize
  The diagrams that drive the renormalization group flow.
  The letters ``$L$'' and ``$R$'' refer to left- and right-handed quarks,
  respectively.
    }
\label{fig:Loops}
}
\end{figure}

Having modeled the four-fermion interaction, we now derive the renormalization
group equations for the couplings.
The diagram that drives the renormalization group flow is the fermion one-loop
diagram and we consider the three cases shown in Figure~\ref{fig:Loops}.
For Diagram (a) in the back-to-back scattering,
we can deduce from the expressions (\ref{eqn:0jsplit}) that
each vertex corresponds to either
\begin{equation}
  iG_{LL}^0 \bar\sigma^{0\dot\alpha\alpha}\bar\sigma^{0\dot\beta\beta}
  \;,\quad\text{or}\quad
  iG_{LL}^j \bar\sigma^{j\dot\alpha\alpha}\bar\sigma^{j\dot\beta\beta}
  \;.
\end{equation}
When the both vertices correspond to $G_{LL}^0$, the diagram yields
\begin{equation}\label{eqn:diagramA00}
  (iG_{LL}^0)^2
  (\bar\sigma^{0\dot\delta\alpha}\bar\sigma^{0\dot\gamma\beta})
  \int\frac{d^4p}{(2\pi)^4}
  \left[
  \frac{ -i (p_\nu - \mu\delta_{\nu,0}) {\sigma^\nu}_{\alpha\dot\alpha}}
  {(p_\lambda - \mu\delta_{\lambda,0})^2}
  \frac{ -i (-p_\eta - \mu\delta_{\eta,0}) {\sigma^\eta}_{\beta\dot\beta}}
  {(-p_\lambda - \mu\delta_{\lambda,0})^2}
  \right]
  (\bar\sigma^{0\dot\alpha\delta}\bar\sigma^{0\dot\beta\gamma})
  \;,
\end{equation}
where we used the quark propagators shown in Appendix~\ref{app:formulas}.
Note that the momentum integral is four dimensional rather than five.
The dynamics of the quarks are restricted to the four dimensional
intersections of the branes and the information about the extra
dimension has been encoded in the form factor.
We decompose the momentum as in Equation~(\ref{eqn:decomposition})
but redefine $\vec p_F$ to include $\vec l_\bot$.
Then near the surface of the large Fermi sphere, 
the vectors $\vec p_F$ and $\vec l_\parallel$ are near parallel and we
also have $E, l_\parallel \ll \mu$ and $p_F\approx\mu$.
Under these approximations together with the use of the $O(3)$-invariance,
the argument of the square bracket in Equation~(\ref{eqn:diagramA00}) becomes
\begin{equation}\label{eq:DAPropPart}
  -\frac{1}{4}\left( -{\sigma^0}_{\alpha\dot\alpha}{\sigma^0}_{\beta\dot\beta}
          +\frac{1}{3}{\sigma^j}_{\alpha\dot\alpha}{\sigma^j}_{\beta\dot\beta}
        \right)
        \frac{1}{ E^2 - l_\parallel^2 }
  \;.
\end{equation}
One can integrate over $E$, either by the contour integral 
or by the Wick rotation,
then $|l_\parallel|$ is integrated from the scale $\Lambda_{UV}$ down to
$\Lambda_{IR}$.
One can also simplify the $\sigma$-matrices (all the necessary formulas
are given in Appendix B of Wess and Bagger~\cite{book:WB}) and 
the expression (\ref{eqn:diagramA00}) becomes
\begin{equation}
  \frac{i\mathcal{N}}{16\pi^3}
  (G_{LL}^0)^2\left(\bar\sigma^{0\dot\delta\delta}\bar\sigma^{0\dot\gamma\gamma}
    -\frac{1}{3}\bar\sigma^{j\dot\delta\delta}\bar\sigma^{j\dot\gamma\gamma}\right)
  t
  \;,
\end{equation}
where we have defined $t:=\ln(\Lambda_{IR}/\Lambda_{UV})$.
This parameter $t$ has the range $(-\infty,0)$ and the lower limit
corresponds to the Fermi surface.
When one of the vertex of Diagram (a) is $G^0$ and the other is $G^j$,
similar procedure yields
\begin{equation}
  \frac{i\mathcal{N}}{16\pi^3}
  (G_{LL}^0G_{LL}^j)
    \left(-2\bar\sigma^{0\dot\delta\delta}\bar\sigma^{0\dot\gamma\gamma}
    +\frac{10}{3}\bar\sigma^{j\dot\delta\delta}\bar\sigma^{j\dot\gamma\gamma}\right)
  t
  \;,
\end{equation}
and when the vertices are both $G^j$, we get
\begin{equation}
  \frac{i\mathcal{N}}{16\pi^3}
  (G_{LL}^j)^2
    \left(5\bar\sigma^{0\dot\delta\delta}\bar\sigma^{0\dot\gamma\gamma}
    -\frac{13}{3}\bar\sigma^{j\dot\delta\delta}\bar\sigma^{j\dot\gamma\gamma}\right)
  t
  \;.
\end{equation}
From those results, we obtain the renormalization group equations
\begin{align}
  \frac{dG_{LL}^0}{dt} &= \frac{\mathcal{N}}{16\pi^3}
                \left\{ -(G_{LL}^0)^2 + 2G_{LL}^0G_{LL}^j - 5(G_{LL}^j)^2 \right\}
  \;,\nonumber\\
  \frac{dG_{LL}^j}{dt} &= \frac{\mathcal{N}}{16\pi^3}
                \left\{ \frac{1}{3}(G_{LL}^0)^2 
                        - \frac{10}{3}G_{LL}^0G_{LL}^j 
                        + \frac{13}{3}(G_{LL}^j)^2 \right\}
  \;.
\end{align}
These equations can be diagonalized to the following forms
\begin{align}
  \frac{d(G_{LL}^0 - 3G_{LL}^j)}{dt}
        &= - \frac{\mathcal{N}}{16\pi^3}(G_{LL}^0 - 3G_{LL}^j)^2
  \;,\\
  \frac{d(G_{LL}^0 + G_{LL}^j)}{dt}
        &= - \frac{\mathcal{N}}{48\pi^3}(G_{LL}^0 + G_{LL}^j)^2
  \;.
\end{align}
One can carry out the same procedure for the back-to-back scattering
of Diagram (b) in Figure~\ref{fig:Loops} and obtain
\begin{align}
  \frac{d(G_{LR}^0 + 3G_{LR}^j)}{dt}
        &= 0
  \;,\\
  \frac{d(G_{LR}^0 - G_{LR}^j)}{dt}
        &= - \frac{\mathcal{N}}{24\pi^3}(G_{L\bar R}^0 - G_{L\bar R}^j)^2
  \;.
\end{align}
Above four renormalization group equations are obtained by 
Evans {\it et al.} \cite{Evans:1998ek}
(but with different form factors).

The Diagram (c) of Figure~\ref{fig:Loops} is similar to the case with (a)
[and not with (b) because $\bar q_R$ has 
an undotted spinor index just as $q_L$].
But we should recall that
we are interested in the forward scattering in this case.
Therefore in the loop, the top left-handed quark propagator
carries the momentum $(E,\,\vec p_F + \vec l_\parallel)$ and the bottom
right-handed one carries $(E,\,-\vec p_F + \vec l_\parallel)$ in the directions
of the arrows in the quark loop.
Then the propagator part of the diagram corresponding to
(\ref{eq:DAPropPart}) of the Diagram (a) is
\begin{equation}
    -\frac{1}{4}\left( {\sigma^0}_{\alpha\dot\alpha}\bar\sigma^{0\dot\beta\beta}
          -\frac{1}{3}{\sigma^j}_{\alpha\dot\alpha}{\bar\sigma^{j\dot\beta\beta}}
        \right)
        \frac{1}{ E^2 - l_\parallel^2 }
  \;.
\end{equation}
This structure is the same as the $LL$-case (a), except the overall sign.
This sign is cancelled by the other one that comes from the
difference in the direction of the quark line.
Thus, apart from the integration range of the scattering angle $\theta$,
the cases (a) and (c) are the same.
So we have for Diagram (c),
\begin{align}
  \frac{d(G_{L \bar R}^0 - 3G_{L \bar R}^j)}{dt}
        &= - \frac{\mathcal{M}}{16\pi^3}(G_{L \bar R}^0 - 3G_{L \bar R}^j)^2
  \;,\\
  \frac{d(G_{L \bar R}^0 + G_{L \bar R}^j)}{dt}
        &= - \frac{\mathcal{M}}{48\pi^3}(G_{L \bar R}^0 + G_{L \bar R}^j)^2
  \;.
\end{align}

Now for the generic form of the renormalization group equation
\begin{equation}
  \frac{dG(t)}{dt} = - K G(t)^2
  \;,
\end{equation}
with some constant $K$, the Landau pole, if exists, is reached at
\begin{equation}
  t_{LP} = - \frac{1}{K G(0)}
  \;.
\end{equation}
Since the range of $t$ is $(-\infty,0)$, the Landau pole exists only
when $K G(0)$ is positive and the larger the factor $K G(0)$ is,
the faster the pole is reached.
We note that because of the constants $\mathcal{N}$ and $\mathcal{M}$ in the
form factors (\ref{eq:FBB}) and (\ref{eq:FFW}), the Landau pole is independent
of those.
From Equations (\ref{eqn:defnG}) and (\ref{eqn:colorFactors}), and also
with the fact that $\mathcal{M}F_{FW}<\mathcal{N}F_{BB}$,
we see that the instability is dominated by the BCS type in
the $LL$, color antisymmetric channel for $N_c\approx 3$.
Note that the color symmetric channel does not have instability
because the interaction in this channel is repulsive.
When $N_c$ is sufficiently large, the $L\bar R$-channel, whose coupling
is proportional to the product of $X(\bullet)$ and $F_{FW}$,
dominates over the other channels because the inequality
$\mathcal{M}F_{FW}<\mathcal{N}F_{BB}$ is compensated by the fact that
$|X(\text{asymm})| < |X(\bullet)|$. 
Thus the DGR-type instability dominates over BCS in this regime.
In our method here, it is not possible to estimate the value of $N_c$
at which the crossover from the BCS- to DGR-type instability occurs,
because we crudely introduced the common infrared cutoff, $m$, to all the
components of the gluon propagator and
we do not have the actual value of $\theta_{UV}$.
See Shuster and Son \cite{Shuster:1999tn} for the estimate of the value $N_c$.

Recall that the form factors grow linearly when the parameters $mL$ and
$\mu/m$ are large and the effective four-fermion couplings become
large accordingly.
We thus expect that the whole analysis breaks down when those parameters
are exceedingly large, and we need to resort to the microscopic analysis, 
{\it i.e.}, the analysis with the fundamental interactions,
to gain insight into the nature of the pathology.
This is the subject of the next analysis.

\subsubsection{Dyson-Schwinger Equations}
\begin{figure}[h]
{
\centerline{\scalebox{0.8}{\includegraphics{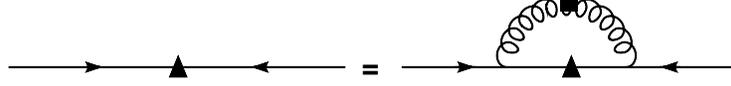}}}
\caption{\footnotesize
  The diagrammatic representation of the Dyson-Schwinger equation
  for the diquark condensate.
  The triangle denotes the gap insertion.
  The square in the gluon propagator represents the screening effect.
    }
\label{fig:DS}
}
\end{figure}

We now turn to the analysis based on the Dyson-Schwinger equations.
The traditional form of the Dyson-Schwinger equation in the diagrammatic
representation is shown in Figure~\ref{fig:DS}.
This method is less intuitive compared to the previous renormalization
group analysis,
but it lets us obtain the actual size of the gap.
This analysis is microscopic which deals with the quarks rather than
the quasi-particles of the effective theory and interactions are
the QCD interactions rather than the effective ones.

As usual, we adopt the Nambu-Gor'kov formalism 
(see, {\it e.g.}, Reference~\cite{Bailin:1983bm}).
In order to introduce the Nambu-Gor'kov basis,
we define the charge conjugate Dirac spinors as
$\psi^C := C\bar\psi^{T}$ and $\bar\psi^C := \psi^TC^T$ where the charge
conjugation matrix $C$ is defined in Appendix~\ref{app:formulas}.
In the Weyl basis, these can be expressed as
\begin{equation}
  \begin{pmatrix}
    \bar q^c_{R\alpha} \\ \bar q_L^{c\,\dot\alpha}
  \end{pmatrix}
  = \begin{pmatrix}
    {(i{\sigma^2}_{\alpha\dot\beta}\,\bar\sigma^{0\dot\beta\gamma})\bar q_{R\gamma}}
    \\
    {(i\bar\sigma^{2\dot\alpha\beta}\,{\sigma^0}_{\beta\dot\gamma})\bar q_L^{\,\ \dot\gamma}}
    \end{pmatrix}
  \;,\quad
  (q_L^{c\,\alpha},q_{R\dot\alpha}^c) = 
  \left(
    {q_L}^\gamma (-i {\sigma^2}_{\gamma\dot\beta}\,\bar\sigma^{0\dot\beta\alpha}),\,
    q_{R\dot\gamma}(-i\bar\sigma^{2\dot\gamma\beta}\,{\sigma^0}_{\beta\dot\alpha})
    \right)
  \;.
\end{equation}
Then we define the Nambu-Gor'kov basis as
\begin{equation}
  \Psi := \frac{1}{\sqrt{2}}
        \left(
   q_{L\alpha},\, {q_R}^{\dot\alpha},\, \bar q^c_{R\beta},\, \bar q_L^{c\,\dot\beta}
        \right)^T
  \;,\quad
  \bar\Psi := \frac{1}{\sqrt{2}}
        \left(
   \bar q_R^{\;\ \alpha},\, \bar q_{L\dot\alpha},\, q_L^{c\,\beta},\, q^c_{R\dot\beta}
        \right)
  \;.
\end{equation}
The advantage of the Nambu-Gor'kov formalism is that we can naturally
include the condensates in the propagator of $\Psi$.
For example, the diquark condensate in the $s$-wave ($LL$ or $RR$ condensate)
is given as
\begin{equation}
  \psi^TC^T\gamma^5\psi = -iq_L^cq_L + iq_R^cq_R
  \;,
\end{equation}
thus the inverse 
propagator that contains this condensate can be written as
\begin{equation}\label{eq:InvNGProp}
  G(p)^{-1} = -i\begin{pmatrix}
             0 & (p_\nu - \mu\delta_{\nu,0})\sigma^\nu & i\bar\Delta_R(p) & 0
             \\
             (p_\nu - \mu\delta_{\nu,0})\bar\sigma^\nu & 0 & 0 & -i\bar\Delta_L(p)
             \\
             i\Delta_L(p) & 0 & 0 & (p_\nu + \mu\delta_{\nu,0})\sigma^\nu
             \\
             0 & -i\Delta_R(p) & (p_\nu + \mu\delta_{\nu,0})\bar\sigma^\nu & 0
            \end{pmatrix}
  \;.
\end{equation}
The $\Delta$-matrices appearing in the inverse propagator are defined as
\begin{equation}\label{eq:DeltaLR}
  \Delta_{L,R}(p) = \Delta_+(p) P_{L,R+}(p) + \Delta_-(p) P_{L,R-}(p)
  \;,
\end{equation}
where $\Delta_\pm(p)$ are the gaps and
the quark (anti-quark) on-shell projectors, $P_{L,R\pm}(p)$, are defined in
Appendix~\ref{app:formulas}.
The ones with the bar can be obtained by replacing $P\to\bar P$ in
the above expression. (We assume the gaps to be real.)
We can invert the matrix (\ref{eq:InvNGProp})
by using the formulas
listed in Appendix~\ref{app:formulas}.%
\footnote{
  Note that the projectors are not invertible, so
  one must use appropriate inversion formulas.
}
If we write
\begin{equation}
  G(p) = \begin{pmatrix}
           G_{11} & G_{12} \\ G_{21} & G_{22}
  			 \end{pmatrix}
  \;,
\end{equation}
we then have
\begin{equation}\label{eq:G21}
  G_{21} = \begin{pmatrix}
  - \frac{\Delta_+ P_{L-}}{p_0^2 - (|\vec p|-\mu)^2 - \Delta_+^2}
  - \frac{\Delta_- P_{L+}}{p_0^2 - (|\vec p|+\mu)^2 - \Delta_-^2}
  & 0
  \\
  0 &
    \frac{\Delta_+ P_{R-}}{p_0^2 - (|\vec p|-\mu)^2 - \Delta_+^2}
  + \frac{\Delta_- P_{R+}}{p_0^2 - (|\vec p|+\mu)^2 - \Delta_-^2}  
  \end{pmatrix}
  \;,
\end{equation}
and other components will not be important in writing down the
Dyson-Schwinger equations.
Notice that if the condensates, $\Delta_\pm$, in the denominators
vanish there are terms
that diverge as the energy is scaled toward the Fermi surface,
{\it i.e.}, $p_0\to 0$ and $|\vec p|\to\mu$.
This is essentially due to the sign structure of the chemical potential
in the matrix (\ref{eq:InvNGProp}) and to the fact that the
$\Delta$-matrices occupy off block-diagonal components.
This infrared divergence is cured by the formation of the condensate and
the $\Delta$ properly behaves as such a condensate.

\begin{figure}[h]
{
\centerline{\scalebox{1.0}{\includegraphics{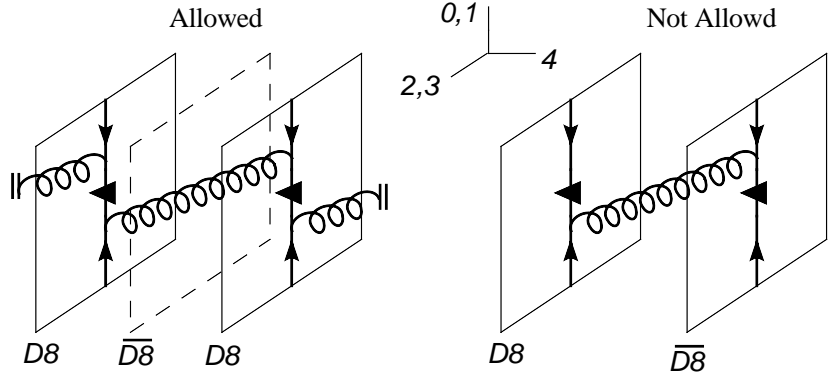}}}
\caption{\footnotesize
  The left diagram represents the right-hand side of the Dyson-Schwinger
  equation (Figure~\ref{fig:DS}) for $LL$-diquark condensate.
  On the right panel is the similar diagram with $LR$-diquark condensate.
  This clearly is not making sense, because the condensate (the triangle)
  is separated and also is not being able to flip the helicity.
 }
\label{fig:DSExtra}
}
\end{figure}

If this were the traditional four-dimensional set up, we could have
introduced the $p$-wave diquark condensate ($LR$-pair)
in the Nambu-Gor'kov propagator.
[Such condensate would have occupied the anti-diagonal slots of the
matrix (\ref{eq:InvNGProp}).]
However, this is not allowed in our theory.
As shown in Figure~\ref{fig:DSExtra},
since $q_L$ and $q_R$ separately live on the D8 and $\Dbar$ branes,
the Dyson-Schwinger equation for such condensate cannot make sense.
In the previous macroscopic renormalization group analysis, 
we encoded all the information
about the extra dimension in the form factors and the left- and right-handed
quarks effectively lived in the same four dimensional spacetime.
However, in this microscopic Dyson-Schwinger analysis, we see that
it is actually not possible for the $LR$-condensate to form.
We emphasize that the condensate is not being energetically suppressed
but simply not possible to form in the brane system of Sakai and Sugimoto.%
\footnote{
 Refs.~\cite{Aharony:2008an,Hashimoto:2008sr} discuss
 the condensation of a gauge invariant
 particle-antiparticle pair across the D8 and $\Dbar$ branes
 at strong coupling.
 Although it is possible that this sort of condensation forms
 in our situation, we note that we are considering the condensation
 of a particle pair which is not gauge invariant and
 at weak coupling.
 Therefore, this possibility is not taken into account in
 our discussion here.
}

For the particle-hole ($L \bar R$ or $\bar L R$) pair,
we might naively introduce the condensate in the diagonal slots of the
matrix in (\ref{eq:InvNGProp}).
However, this represents the introduction of the usual chiral
condensate, that is, the pair of particle and anti-particle, and not
the desired particle-hole pair.
One can verify in this case that the infrared divergence near the
Fermi surface is not present.
Thus, the Nambu-Gor'kov basis is not suitable for introducing the
particle-hole chiral condensate.
Instead, we propose to use the doubled basis,
\begin{equation}
  \frac{1}{\sqrt 2}\left(
    q_{L\alpha},\, {q_R}^{\dot\alpha},\, 
    q_{L\alpha},\, {q_R}^{\dot\alpha}
  \right)^T
  \;,
\end{equation}
and consider the inverse propagator of the form (\ref{eq:InvNGProp})
with the replacements, $i\Delta_L\to\Sigma_L$, $-i\Delta_R\to\Sigma_R$, 
$-i\bar\Delta_L\to\bar\Sigma_L$ and $i\bar\Delta_R\to\bar\Sigma_R$,
where the $\Sigma$-matrices are similarly defined as for the
$\Delta$-matrices.
The spinors of the second set have the chemical potentials in their
kinetic terms with opposite sign from the first set.
This effectively introduces the hole degrees of freedom.
In this case, the propagator has the infrared divergence
near the Fermi surface and the $\Sigma$-condensate provides the cutoff.
Thus we properly have the interpretation that the condensate is the
particle-hole pair near the Fermi surface.
We note that since the condensate is formed out of the spinors with
the dotted- or undotted-index pair, and not the mixed one,
the condensate lives either on D8 or $\Dbar$ branes and not
across them.
%

We must also consider the gluon propagator which is the other ingredient
of the Dyson-Schwinger equation.
Unlike previous macroscopic treatment, we properly take into account
of the perturbatively computable screening effect.
The most general form of the $O(3)$-invariant gluon propagator is
\begin{equation}
  D_{\mu\nu}(p,n) =
    \frac{P^T_{\mu\nu}(p)}{p^2 + (2\pi n/L)^2 + G_s(p)}
    +
    \frac{P^L_{\mu\nu}(p)}{p^2 + (2\pi n/L)^2 + F_s(p)}
  \;,
\end{equation}
where $F_s$ and $G_s$ are the electric and magnetic screenings, respectively,
and the projectors are defined as
\begin{equation}
  P^T_{ij} (p) = \eta_{ij} - \frac{p_i p_j}{|\vec p|^2}
  \;,\;
  P^T_{00} (p) = 0 = P^T_{01} (p)
  \;,\;
  P^L_{\mu\nu} (p) = \eta_{\mu\nu} - \frac{p_\mu p_\nu}{p^2} 
  - P^T_{\mu\nu}
  \;.
\end{equation}
We have dropped the term with the gauge fixing parameter in the
propagator.
This term has been verified not to contribute to the gap, the solution
to the Dyson-Schwinger equation, at very large chemical potential
\cite{Rajagopal:2000rs}.

As explained in Section~\ref{subsec:qualitative}, the finite density
screening effect is given by the diagram shown in
Figure~\ref{fig:quarkLoop}, at one-loop level.
Since the quark loop stays on the D8 or $\Dbar$ branes, there is no extra
dimensional effect on the screening,
and the gluons with nonzero
momentum in the $x_4$-direction, such as the case shown in
Figure~\ref{fig:DSExtra}, do not have the screening.
%
%
We thus have the standard expressions \cite{book:LeBellac}
\begin{align}
  F_s(p) &= 2m_D^2 \frac{p^2}{|\vec p|^2}
  			\left[
  			  1 - \frac{p_0}{|\vec p|} Q_0 \left(\frac{p_0}{|\vec p|}\right)
  			\right] \delta_{n,0}
  \;,\nonumber\\
  G_s(p) &= m_D^2 \frac{p_0}{|\vec p|}
        \left[
          \left\{ 1 - \left(\frac{p_0}{|\vec p|}\right)^2 \right\}
          Q_0 \left(\frac{p_0}{|\vec p|}\right)
          + \frac{p_0}{|\vec p|}
        \right]\delta_{n,0}
  \;,\nonumber\\
  Q_0 (x) &=
    \frac{1}{2} \ln \left|\frac{1+x}{1-x}\right|
    -i\frac{\pi}{2} \theta (1-x^2)
  \;,\quad
  m_D^2 = \frac{1}{4\pi^2} N_f g^2 \mu^2
  \;,
\end{align}
where $\theta$ is the Heaviside function and $\delta_{n,0}$ signifies
that the screening is effective only for the gluons with $n=0$.
\\

We can now write down the Dyson-Schwinger equation.
We start with the equation for the $\Delta$-condensate.
At high density (weak coupling), we can approximate the quark-gluon
vertex with the bare ones and we have
\begin{equation}\label{eq:firstDS}
  G(k)^{-1} - G_0 (k)^{-1} =
  	-i	g^2 X(D) \sum_n \int \frac{d^4 p}{(2\pi)^4}
  		\Gamma^\mu G(p)\Gamma^\nu D_{\mu\nu}(q,n)
  \;.
\end{equation}
We have included the color factor $X(D)$ as in
Equations~(\ref{eqn:colorFactors}) and the representation $D$ is 
either symmetric or antisymmetric depending on the channel that
$G$ is in.%
\footnote{
  The easiest way to see how this color factor comes in is to note
  that
  \begin{equation}
    \sum_a T^\alpha_{ab}T^\alpha_{cd} = 
   \frac{1}{2}X(\text{symm})(\delta_{ab}\delta_{cd}+\delta_{ad}\delta_{bc})
 +\frac{1}{2}X(\text{asymm})(\delta_{ab}\delta_{cd}-\delta_{ad}\delta_{bc})
  \;.
  \end{equation}
}
Also we have defined $G_0$ as $G$ without the condensates, $q:=k-p$,
and $\Gamma^\mu$ are
\begin{equation}
  \Gamma^\mu := \begin{pmatrix}
    0 & \sigma^\mu & 0 & 0
    \\
    \bar\sigma^\mu & 0 & 0 & 0
    \\
    0 & 0 & 0 & -\sigma^\mu
    \\
    0 & 0 & -\bar\sigma^\mu & 0
  \end{pmatrix}
  \;.
\end{equation}
We can use (\ref{eq:InvNGProp}),
(\ref{eq:G21}) and (\ref{eq:firstDS}) to obtain the gap equation
\begin{align}
  \Delta_\pm(k) = &
  ig^2 X(D) \sum_n\int \frac{d^4p}{(2\pi)^4}
  \nonumber\\
  &\tr \bigg[
    \sigma^\mu \bigg(
     \frac{\Delta_+(p) P_{R-}(p)}{p_0^2 - (|\vec p|-\mu)^2 - \Delta_+^2}
     + \frac{\Delta_-(p) P_{R+}(p)}{p_0^2 - (|\vec p|+\mu)^2 - \Delta_-^2}  
    \bigg)
    \bar\sigma^\nu P_{L\pm}(k)
  \bigg] D_{\mu\nu}(q,n)
  \;,
\end{align}
where the trace is over the spinor indices and is taken to project
out $\Delta_\pm$ from the $\Delta_L$-matrix in Equation~(\ref{eq:DeltaLR}).
After some algebra, one obtains
\begin{align}
  \Delta_{+}(k_0) \approx &
  -ig^2 X(D) \sum_n\int \frac{d^4p}{(2\pi)^4}
  \nonumber\\
   \bigg[
    & \frac{\Delta_+ (p_0)}{p_0^2 - (|\vec p|-\mu)^2 - \Delta_+(p_0)^2}
     \bigg(
       \frac{1-(\hat p \cdot \hat q)(\hat k \cdot \hat q)}
       {q^2 + (2\pi n/L)^2 + G_s(q)}
       + \frac{\frac{1}{2}+\frac{1}{2}\hat p \cdot \hat k}
       {q^2 + (2\pi n/L)^2 + F_s(q)}
     \bigg)
  \nonumber\\
  + & \frac{\Delta_- (p_0)}{p_0^2 - (|\vec p|+\mu)^2 - \Delta_-(p_0)^2}
     \bigg(
       \frac{1+(\hat p \cdot \hat q)(\hat k \cdot \hat q)}
       {q^2 + (2\pi n/L)^2 + G_s(q)}
       + \frac{\frac{1}{2}-\frac{1}{2}\hat p \cdot \hat k}
       {q^2 + (2\pi n/L)^2 + F_s(q)}
     \bigg)
   \bigg]
  \;,
\end{align}
where we have assumed that the gaps are functions only of $k_0$ or
$p_0$.
In deriving this equation, we have adopted the approximation,
$q_0 \ll |\vec q| \approx \mu$, so that 
$P^L_{\mu\nu} \approx \eta_{\mu\nu}\delta_{\mu,0}\delta_{\nu,0}$.
The equation for $\Delta_-(k_0)$ is the same except that the two terms
in the round brackets are exchanged.
Note that only the first term in the equation of $\Delta_+(k_0)$ has
the near-Fermi-surface (infrared) divergence that is being cured by
the formation of the condensate.
Thus to the first approximation in large $\mu$, we can neglect the
second term.
Similar observation in the equation of $\Delta_-(k_0)$ results in the
conclusion that this gap does not form at near the Fermi surface.

Now as before, we set $\vec p = \vec p_F + \vec l_\parallel$,
and approximate $|\vec p_F|\approx\mu$, 
$\hat p_F \cdot \hat l_\parallel \approx 1$,
$|\vec p| \approx \mu + l_\parallel$
and $q^2 \approx 2\mu^2 (1-\hat p \cdot \hat k)$.
Then the integration measure takes the form
$\mu^2dp_0dl_\parallel d\cos\theta d\phi$ with
$\cos\theta := \hat p \cdot \hat k$.
Since the integral is dominated by the region $\theta\approx 0$,
we further approximate that $\hat p \cdot \hat k \approx 1$
and $\hat p \cdot \hat q \approx 0 \approx \hat k \cdot \hat q$
in the numerators of the integral, but not in the denominators.
We Wick rotate $p_0\to ip_0$ and integrate over $l_\parallel$ and
$\phi$.
The gap equation now takes the form
\begin{align}\label{eq:afterApprox}
  \Delta_+(k_0) &\approx 
  -\frac{g^2}{8\pi^2}X(D)\sum_n\int dp_0 d\cos\theta
   \bigg(
      \frac{1}{1 - \cos\theta + (1/2)\{2\pi n/(\mu L)\}^2
       + G_s(q)/(2\mu^2)}
  \nonumber\\
     & +
      \frac{1}{1 - \cos\theta + (1/2)\{2\pi n/(\mu L)\}^2
       + F_s(q)/(2\mu^2)}     
    \bigg)
    \frac{\Delta_+(p_0)}{\sqrt{p_0^2+\Delta_+(p_0)^2}}
  \;,
\end{align}
with the approximate form of the screenings
\begin{equation}\label{eq:approxSreening}
  F_s(q) \approx 2m_D^2 \, \delta_{n,0}
  \;,\quad
  G_s(q) \approx \frac{\pi}{2}m_D^2 \frac{q_0}{|\vec q|} \, \delta_{n,0}
  \;.
\end{equation}
From this equation, it is clear that for the symmetric channel
$D=\text{symm}$, we only have the trivial solution $\Delta_+=0$.%
\footnote{
  If the flavor structure is included, this conclusion gets slightly more
  complicated.
  However, the fact that the antisymmetric channel dominates over the
  symmetric one does not change. See, for instance,
  Reference~\cite{Shovkovy:1999mr}.
}
We thus consider the antisymmetric channel from now on.

In Equation~(\ref{eq:afterApprox}), the sum over $n$ and integral
over $\theta$ can be carried out in a straightforward manner and
yields
\begin{equation}\label{eq:son}
  \Delta_+(k_0) \approx \frac{g^2}{12\pi^2}\frac{N_c+1}{2N_c}
  \int dp_0 \ln \left( \frac{\Lambda}{|k_0 - p_0|} \right)
  \frac{\Delta_+(p_0)}{\sqrt{p_0^2+\Delta_+(p_0)^2}}
  \;,
\end{equation}
where we have defined
\begin{equation}
  \Lambda := 2^{10}\sqrt 2\pi^4 N_f^{-5/2}g^{-5}\mu
  					\{\sinh(\mu L)/(\mu L)\}^4
  \;,
\end{equation}
and the part, $\{\cdots\}^4$, is the contribution from
the sum over $n\neq 0$.
The factor $|k_0 - p_0|$ appearing in the logarithm comes from
the Landau damping of $G_s(q)$ as in (\ref{eq:approxSreening})
and this effect was first discussed by Son~\cite{Son:1998uk}.
We can follow Appendix B of Reference~\cite{Son:1998uk}
to solve this equation and obtain
\begin{equation}\label{eq:LLgap}
  \Delta_+ (k_0) = \Delta_0 \sin\left( 
  			\sqrt{\frac{12\pi^2}{g^2}\frac{2N_c}{N_c+1}}
  			\ln\frac{\Lambda}{k_0}
  			\right)
  \;,\quad\text{with}\quad
  \Delta_0 = \Lambda\exp\left[
  			  -\frac{\pi}{2}\sqrt{\frac{12\pi^2}{g^2}\frac{2N_c}{N_c+1}}
  			\right]
  \;.
\end{equation}

Let us comment on this result.
We first note that the gap vanishes in the 't Hooft limit.
This is consistent with what we have concluded in
the renormalization group analysis.
Now, when $\mu L \ll 1$, we have $\sinh(\mu L)/\mu L \approx 1$, so the
extra dimensional effect in $\Lambda$ disappears and
the resulting expression for the gap coincide with the QCD result
for $N_c=3$. (See, for example, Reference~\cite{Rajagopal:2000wf}.)
In the opposite limit where $\mu L\to\infty$, 
we have $\sinh(\mu L)\to \exp(\mu L)/2$, so the extra
dimensional effect contributes heavily and the gap grows with the
parameter $\mu L$.
As one can observe in Equation~(\ref{eq:afterApprox}), this is because
the infrared effect of the terms with $n \neq 0$ comes to be comparable
to that of the $n=0$ term.
From Equation~(\ref{eq:LLgap}), we see that at $\mu L \sim 1/g$,
the gap is much larger than the size of the Fermi sphere itself
and such a solution cannot be accepted, for the dynamics is no longer
taking place near the Fermi surface.
We therefore conclude that when $\mu L \sim 1/g$, the gap does not
form and the ground state simply is described by the Fermi liquid.
\\

We now turn to the $\Sigma$-condensate, $\chi$DW.
The computation is almost identical to the previous case and we arrive at the
equation similar to Equation~(\ref{eq:afterApprox}) with the replacements
$\Delta_+\to\Sigma_+$ and $D\to\bullet\,$.
There are, however, a few differences.
The most important one is the range of
the integration parameter $\theta$.
This is restricted to the near infrared region, {\it i.e.}, $\theta\approx 0$,
because this is a forward scattering and the exchanged gluon should
not be harder than the size of the gap
or the momentum carried by the propagator $G(k)$.
When the angle is set to small value, the $l_\parallel$ component
is about $\mu(1-\cos\theta)$, and the propagator carries the momentum
approximately $\sqrt{p_0^2+\Sigma_+^2}$.
We thus require
\begin{equation}\label{eq:thetaUpper}
  \mu(1-\cos\theta) \leq \sqrt{p_0^2+\Sigma_+^2}
  \;,
\end{equation}
and this inequality sets the upper limit on $\theta$.
Now because of this kinematic restriction,
when $N_c\approx 3$, the $\Delta$-condensate dominates over
the $\Sigma$-condensate.
Therefore, in the following, we consider the 't Hooft limit
(with small $\lambda$).
In this limit, the screening $F_s$ and $G_s$ is $1/N_c$-suppressed,
so we drop the screening terms from the gap equation.
In the absence of the screening, the approximation,
$q^2 \approx 2\mu^2 (1-\cos\theta)$, has infrared problem when $n=0$.
This means that $q_0^2$ cannot be neglected in this case and
we must use 
$q^2 \approx 2\mu^2 (1-\cos\theta) + |k_0-p_0|^2\delta_{n,0}$.
We thus have the gap equation for the $\Sigma$-condensate
\begin{align}
  \Sigma_+(k_0) =& \frac{g^2}{8\pi^2}\frac{N_c^2-1}{2N_c}
  \sum_n\int dp_0 d\cos\theta
  \nonumber\\
  &\frac{2}{1-\cos\theta+|k_0-p_0|^2/(2\mu^2)\delta_{n,0}
    + (1/2)\{2\pi n/(\mu L)\}^2}
  \frac{\Sigma_+(p_0)}{\sqrt{p_0^2+\Sigma_+(p_0)^2}}
  \;,
\end{align}
where the integration range of $\theta$ is restricted as mentioned above.

Let us first consider the case with $n=0$.
In this case, we can carry out the $\cos\theta$ integral with the
restriction (\ref{eq:thetaUpper}) and obtain
\begin{equation}\label{eq:sigmaGap}
  \Sigma_+(k_0) \approx
  \frac{g^2}{8\pi^2}\frac{N_c^2-1}{N_c} \int dp_0
  \ln \left(
      \frac{2\mu\sqrt{p_0^2+\Sigma_+(p_0)^2}}{|k_0-p_0|^2}
    \right)
    \frac{\Sigma_+(p_0)}{\sqrt{p_0^2+\Sigma_+(p_0)^2}}
   \;.
\end{equation}
We can again solve this equation following Son \cite{Son:1998uk}
(also see \cite{Park:1999bz}).
In this case, the calculation is slightly different from
Reference~\cite{Son:1998uk}, so it is shown in Appendix~\ref{app:solveGap}.
The result is
\begin{equation}\label{eq:resultSigma}
  \Sigma_+(k_0)=
  \Sigma_0\cos\left(\sqrt{\frac{g^2}{4\pi^2}\frac{N_c^2-1}{N_c}}\ln\frac{2\mu}{k_0}\right)
  \;,\quad\text{with}\quad
  \Sigma_0 = 
  2\mu\exp\left[-\pi\sqrt{\frac{4\pi^2}{g^2}\frac{N_c}{N_c^2-1}}\right]
  \;.
\end{equation}
This result agrees with Reference~\cite{Deryagin:1992rw}.

When we include the terms with $n\neq 0$, the
logarithm term in Equation~(\ref{eq:sigmaGap}) gets augmented as
\begin{equation}\label{eq:logLog}
  \ln\frac{2\mu\sqrt{p_0^2+\Sigma_+(p_0)^2}}{|k_0-p_0|^2}
  +
  4\ln\frac{\sinh(\mu L\epsilon^{1/2})}{\mu L\epsilon^{1/2}}
  \;,
\end{equation}
where $\epsilon := \sqrt{p_0^2+\Sigma_+^2}/\mu$.
If $\mu L$ is not too large compared to $\epsilon^{1/2}$,
then the second term in the above expression is small and the
result (\ref{eq:resultSigma}) does not change.
However, when $\mu L$ is so large that the second term yields dominant
contribution $\sim\mu L\epsilon^{1/2}$, the integrand of the gap equation
(\ref{eq:sigmaGap}) takes the from proportional to $1/(p_0^2+\Sigma_+^2)^{1/4}$,
which is free of the infrared divergence even without the condensate $\Sigma$.
This implies that the gap does not exist.
To roughly estimate the value of $\mu L$ at which the crossover occurs,
we approximate the first term in Equation~(\ref{eq:logLog}) as
$\ln (2\mu/\Sigma_0)$ and the second term as
$\mu L(\Sigma_0/\mu)^{1/2}$.
Then we see that the second term becomes important when
\begin{equation}
  \mu L \gtrsim (\mu/\Sigma_0)^{1/2}\ln(\mu/\Sigma_0) \approx
  e^{1/\sqrt{\lambda}}/\sqrt{\lambda}
  \;.
\end{equation}

\section{Comments on the Strong Coupling Regime}
\label{sec:discussions}
We have addressed the possibilities of the color superconductivity
and the chiral density waves in the Sakai-Sugimoto model at finite
density and explicitly carried out the computations
in the weak coupling limit.
As was stated in the introduction, the ultimate goal of this
investigation is
to obtain the quantitative behavior of the model in the weak and
strong coupling regions, make qualitative comparison of the phase diagrams
and gain insight into the QCD phase diagram which is still unsettled.
We therefore comment on the strong coupling gravity background analysis
of the model at finite density.

As suggested by Sakai and Sugimoto \cite{Sakai:2004cn},
the strong coupling analysis is done by taking the gravity background
limit of the D4-branes while treating the D8-branes as probes,
then the DBI-action of the probes are studied to obtain the spectrum of
the low energy excitations at strong coupling.
The generalization to the finite density has been discussed in
References~\cite{Kim:2006gp,Horigome:2006xu,Parnachev:2006ev}.
The phase diagram of the model in the space of temperature
and chemical potential
has been obtained by Horigome and Tanii \cite{Horigome:2006xu}.
Let us briefly review their results.
In the previous section, we have placed the D8- and $\Dbar$-branes
at the antipodal points of the compactified circle.
However, this is not necessary and in this gravity background analysis,
{\it the compactification radius is set to $R$ and the distance
between the branes to $L$} with the range $0<L\leq\pi R$.

\begin{figure}[h]
{
\centerline{\scalebox{0.6}{\includegraphics{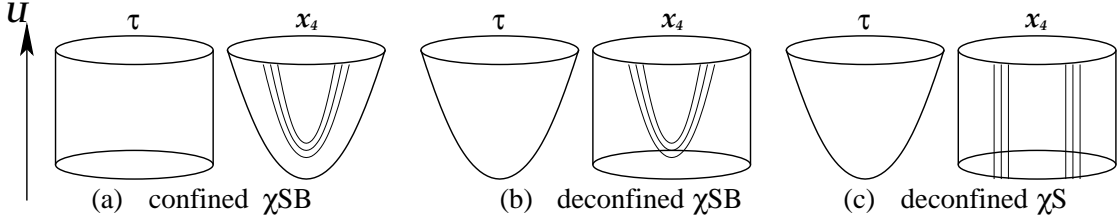}}}
\caption{\footnotesize
  The spacetime configurations of the three phases.
  The vertical axis $U$ is the radial direction in 5,6,7,8 and 9
  directions.
  The thinner lines represent the stacks of D8 and $\Dbar$ branes.
  Diagram (a) is the low temperature confined phase. The chiral symmetry
  is broken in this phase.
  Diagram (b) is the high temperature deconfined phase and with
  $\chi$SB.
  Diagram (c) is also the high temperature deconfined phase but with
  chiral symmetry restored.
    }
\label{fig:threePhases}
}
\end{figure}

At finite temperature, in addition to the $x_4$-direction,
the Euclidean time direction, $\tau$, is also compactified and the
period of the time circle is identified with the inverse temperature.
Horigome and Tanii consider the
three known phases, first discovered by 
Aharony {\it et al.} \cite{Aharony:2006da} at zero density,
and the spacetime configurations
for the phases are illustrated in Figure~\ref{fig:threePhases}.
In the figure, the vertical $U$-axis represents the radial direction
in 5,6,7,8 and 9 directions.
The change in the background geometry of the phases from (a) to (b,c)
is interpreted as the confinement/deconfinement phase transition
by Aharony {\it et al}.
The thinner lines and curves in the diagrams represent the D8 or
$\Dbar$ branes and the change in the configuration from Diagram (b)
to (c) is interpreted as the chiral symmetry restoration.

\begin{figure}[h]
{
\centerline{\scalebox{0.6}{\includegraphics{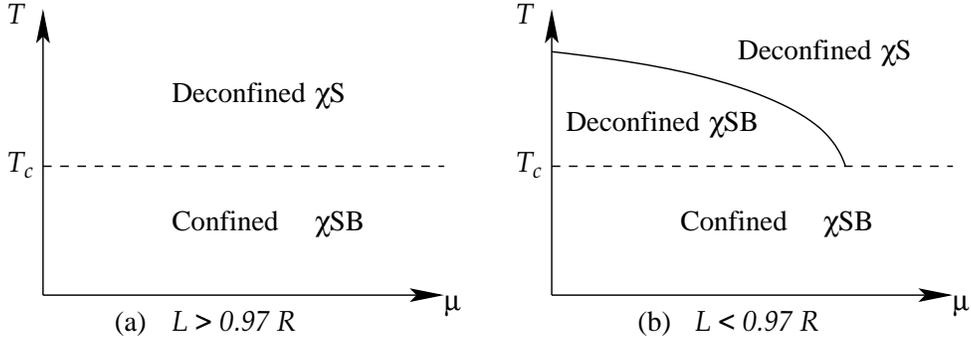}}}
\caption{\footnotesize
  The schematic phase diagram obtained by Horigome and Tanii.
  The temperature $T_c$ denotes the confinement/deconfinement
  phase transition temperature.
  The inter-D8$\Dbar$ distance $L\simeq 0.97R$ is the critical value
  where the deconfined phase with $\chi$SB exists.
    }
\label{fig:HTPhaseD}
}
\end{figure}

The on-shell DBI actions of the D8-branes with nonzero chemical potential
have been obtained by Horigome and Tanii for each phase.
For the configurations with the {\it smooth} U-shaped D8-branes, 
that is, for the diagrams (a) and (b) of Figure~\ref{fig:threePhases},
they have found that the chemical potential must be constant along the
radial $U$-direction and the actions for those configurations are
independent of the chemical potential.
Only for the parallel D8-brane configuration of Diagram (c) has
the non-trivial dependence on the chemical potential in its action.
By comparing the on-shell actions at the various values of the temperature
and the chemical potential, they determined the phase diagram which is
schematically shown in Figure~\ref{fig:HTPhaseD}.%
\footnote{
  In Reference~\cite{Horigome:2006xu}, the temperature and chemical potential
  are measured in different units.
  If measured in the same units, the chemical potential is asymptotically
  larger than the temperature by the factor of the 't Hooft coupling
  $\lambda$.
}

The confinement/deconfinement phase transition line at $T=T_c$ is determined
by the D4-background geometry and the D4 difference action in those
phases scales as $N_c^2$ \cite{Aharony:2006da}.
This is expected to completely dominate the D8 probe actions, which scale as
$N_c N_f$, so the confinement/deconfinement phase transition line is not
affected by the value of the chemical potential.%
\footnote{
  In the original version of this e-print, we overlooked the dominance of
  the D4 action and compared only the D8 actions in the confined and deconfined
  phases, which was incorrect.
}
Therefore, Horigome and Tanii assume that the phase below
the confinement/deconfinement line is in the configuration (a) of
Figure~\ref{fig:threePhases}, and compare the D8 actions of the configurations
(b) and (c) in the deconfined D4-background geometry to obtain the phase
structure above the confinement/deconfinement phase transition line.
This is why the $\chi$S/$\chi$SB phase transition line of the right
panel in Figure~\ref{fig:HTPhaseD} is terminated at $T=T_c$.
When the inter-D8$\Dbar$ distance $L$ is larger than $0.97 R$,
it has been shown by Aharony {\it et al.} \cite{Aharony:2006da}
that the phase represented in Diagram (b) of Figure~\ref{fig:threePhases}
does not exist.
Therefore, the phase diagram becomes rather structureless, as shown in
the left panel of Figure~\ref{fig:HTPhaseD}.
\\

\begin{figure}[h]
{
\centerline{\scalebox{0.3}{\includegraphics{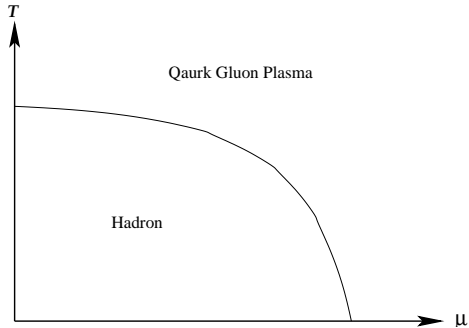}}}
\caption{\footnotesize
  Expected phase diagram of the simplified QCD where only the hadronic and
  quark-gluon plasma phases are assumed to exist.
    }
\label{fig:simplified}
}
\end{figure}

The analysis of Horigome and Tanii assumes only three
possible phases and this is similar to the simplified picture of QCD where
one assumes only hadronic and quark-gluon plasma phases.
In the latter, it is expected that when the chemical potential is raised
several times $\Lambda_{QCD}$ (at low temperature), 
the hadrons start to overlap and the quarks are shared by many hadrons.
This implies the change in the degrees of freedom and since only the two
phases are assumed, there should be a phase transition.
Stated differently, in this simplified picture of QCD,
we expect that a phase transition to occur even at zero temperature.
See Figure~\ref{fig:simplified}.%
\footnote{
  It is interesting to observe that the weakly coupled large $N_c$
  $\mathcal{N}=4$ super-Yang-Mills theory on three sphere and Type IIB
  supergravity on AdS$_5\times S^5$ both have qualitatively the same
  phase structure as Figure~\ref{fig:simplified}
  \cite{Yamada:2006rx,Chamblin:1999tk,Cvetic:1999ne}.
  In these cases, however, the chemical potential is conjugate to
  the $U(1)$ subgroup of $SU(4)$ $R$-symmetry.
}
Though the physical setups are similar, we see that the phase diagrams in
Figures~\ref{fig:HTPhaseD} and \ref{fig:simplified} are qualitatively
different.
We suspect that the discrepancy stems from the large $N_c$ limit and
the probe approximation in the Sakai-Sugimoto model.
As we have explained, the confinement/deconfinement phase transition is
determined by the D4 actions because the difference action scales with $N_c^2$
and dominates over the D8 actions (with the chemical potential).
It is unlikely that this picture can change in the probe
approximation.
Therefore, in order to see modifications in the confinement/deconfinement
phase transition line, one should take into account of the D8-brane back
reaction to the geometry.

It is also clear that we do not observe the color superconductivity or
chiral density waves in the strong coupling analysis of Horigome and Tanii
because these possibilities are simply not considered.
To explore those exotic phases, one must come up with the corresponding
stringy pictures of the branes and strings, and see if they are energetically
preferred at any point in the $\mu$-$T$ parameter space.
Such stringy pictures of quark matter are, so far, not clear to us.%
\footnote{
 The stringy configurations of baryons have been suggested
 in Refs.~\cite{Bergman:2007wp,Rozali:2007rx}
 in which the nuclear matter phase has been
 discovered in the phase diagram.
}
Nevertheless, we make some general remarks in this direction.
First, we must remember that the relevant degrees of freedom at high
density are particles and holes while the anti-particles are buried
deep in the Dirac sea.
Thus, if the U-shape configuration of the D8-branes describes the mesons,
which are the pairs of particle and anti-particle, then we expect
the configuration of the superconductor or chiral density waves to be
different from the U-shape configuration because they are
the pairs of particles and holes.
Secondly, the Fermi sphere plays the essential role in high density QCD.
Thus, the Fermi sphere must be encoded in the holographic picture somehow.
Also,
the holography is in the 't Hooft limit under which we do not expect
non-gauge invariant quantities to survive.
Therefore, it is likely that we do not observe
the color superconducting phase and the relevant phase probably is the chiral
density waves in the holographic theories.

Although Shuster and Son \cite{Shuster:1999tn} settled that
the color superconductivity dominates over the chiral density waves
in the high density weakly coupled QCD (with $N_c=3$),
it still is not known if this observation persists at the medium
density strong coupling region.
The Sakai-Sugimoto model in the usual 't Hooft limit is not likely
to be able to address this problem but it is interesting to examine
this in the limit with $N_f/N_c$ fixed.
In this case, we can expect the effect of the color-flavor locking to
be significant and also the competition between the D4 and D8 actions
becomes non-trivial.

Finally, as we have mentioned in the introduction, the rich structure
of the QCD phase diagram is partly due to the quark masses.
As the authors of Reference~\cite{Sugimoto:2004mh} have already
noted, the inclusion of the quark masses involves the tachyon
that comes from the string stretching between the D8 and $\Dbar$ branes.
This subject is being actively studied
(for example, see Refs.\cite{Bergman:2007pm,Dhar:2007bz,Dhar:2008um,Aharony:2008an,Hashimoto:2008sr})
and the resulting phase diagram is yet to be seen.

{\it If} the finite density holographic model
turns out to capture the aspects of QCD,
it would be very interesting because we can explore the region of
the QCD phase diagram where the perturbative nor the numerical analysis
is available.


\section*{Acknowledgments}
I would like to thank the members of Racah Institute of Physics
who provided valuable questions and comments.
I benefited from Oren Bergman and Jacob Sonnenschein with regard to
the strongly coupled aspects of the theory.
I also would like to thank Andrei Kryjevski for educating me
high density QCD in year 2004.
This work was supported by the Golda Meir Post-Doctoral fellowship.

\bigskip
\bigskip

\appendix

\section{Some Formulas}\label{app:formulas}
We adopt the convention of Wess and Bagger~\cite{book:WB},
except the definition of the Dirac spinor as in Equation~(\ref{eq:Dirac}).
Many useful formulas can be found in Appendix B of the reference.

The propagators of the left- and right-handed quarks are respectively
given as
\begin{equation}
  \frac{ -i (p_\nu - \mu\delta_{\nu,0}) {\sigma^\nu}_{\alpha\dot\alpha}}
  {(p_\lambda - \mu\delta_{\lambda,0})^2}
  \;,\quad\text{and}\quad
  \frac{ -i (p_\nu - \mu\delta_{\nu,0}) \bar\sigma^{\nu\dot\alpha\alpha}}
  {(p_\lambda - \mu\delta_{\lambda,0})^2}
  \;.
\end{equation}

We define the charge conjugation matrix
\begin{equation}
  C := 	\begin{pmatrix}
	  i\sigma^2\bar\sigma^0 & 0 
          \\ 
          0 & 	  i\bar\sigma^2\sigma^0
	\end{pmatrix}
  \;,
\end{equation}
which satisfies $C^{-1}\gamma^\nu C = -\gamma^{\nu T}$ and
$C^{-1} = C^{T} = - C$.

The quark and anti-quark on-shell projectors are defined as
\begin{align}
  P_{L\pm}(p) &:= \frac{1}{2}\left( 1 \pm \sigma^0\bar\sigma^j\hat p_j \right)
  \;,\quad
  P_{R\pm}(p) := \frac{1}{2}\left( 1 \pm \bar\sigma^0\sigma^j\hat p_j \right)
  \;,\nonumber\\
  \bar P_{L\pm}(p) &:= \frac{1}{2}\left( 1 \pm \hat p_j\bar\sigma^j\sigma^o \right)
  \;,\quad
  \bar P_{R\pm}(p) := \frac{1}{2}\left( 1 \pm \hat p_j\sigma^j\bar\sigma^0 \right)
  \;,
\end{align}
where $\hat p := \vec p/|\vec p|$.
Notice that $P_{L\pm} = \bar P_{R\mp}$ and $P_{R\pm} = \bar P_{L\mp}$.

We list the useful formulas for inverting the inverse of the Nambu-Gor'kov
propagator.
For $n\times n$ matrices $A$, $B$, $C$ and $D$, we have
\begin{equation}
  \begin{pmatrix}
    A & B \\ C & D
  \end{pmatrix}^{-1}
  =
  \begin{pmatrix}
    A^{-1} + A^{-1}BS_A^{-1}CA^{-1} & -A^{-1}BS_A^{-1}
    \\
    -S_A^{-1}CA^{-1} & S_A^{-1}
  \end{pmatrix}
  \;,
\end{equation}
provided that the matrices $A$ and $S_A := D-CA^{-1}B$ are invertible.
We also have
\begin{equation}
  \begin{pmatrix}
    A & B \\ C & D
  \end{pmatrix}^{-1}
  =
  \begin{pmatrix}
    -C^{-1}DS_C^{-1} & C^{-1}+C^{-1}DS_C^{-1}AC^{-1}
    \\
    S_C^{-1} & -S_C^{-1}AC^{-1}
  \end{pmatrix}
  \;,  
\end{equation}
provided that $C$ and $S_C := B-AC^{-1}D$ are invertible.
Other convenient formulas are the following.
\begin{align}
  (p_\nu - \mu\delta_{\nu,0}) \sigma^\nu
  &= (p_0 + |\vec p| - \mu) \sigma^0 P_{R+}
  + (p_0 - |\vec p| - \mu) \sigma^0 P_{R-}
  \;,\nonumber\\
  (p_\nu - \mu\delta_{\nu,0}) \bar\sigma^\nu
  &= (p_0 + |\vec p| - \mu) \bar\sigma^0 P_{L+}
  + (p_0 - |\vec p| - \mu) \bar\sigma^0 P_{L-}
  \;.
\end{align}
\begin{align}
  \{(p_\nu - \mu\delta_{\nu,0}) \sigma^\nu\}^{-1}
  &= \frac{\bar\sigma^0 P_{L+}}{p_0 - |\vec p| - \mu}
   + \frac{\bar\sigma^0 P_{L-}}{p_0 + |\vec p| - \mu}
  \;,\nonumber\\
  \{(p_\nu - \mu\delta_{\nu,0}) \bar\sigma^\nu\}^{-1}
  &= \frac{\sigma^0 P_{R+}}{p_0 - |\vec p| - \mu}
   + \frac{\sigma^0 P_{R-}}{p_0 + |\vec p| - \mu}
  \;. 
\end{align}
\begin{align}
  P_{L\pm}\,\sigma^0P_{R\pm} &= 0
  \;,\quad
  P_{L\pm}\,\sigma^0P_{R\mp} = \sigma^0P_{R\mp}
  \;,\nonumber\\
  P_{R\pm}\,\bar\sigma^0P_{L\pm} &= 0
  \;,\quad
  P_{R\pm}\,\bar\sigma^0P_{L\mp} = \bar\sigma^0P_{L\mp}
  \;.
\end{align}

\section{Solving Gap Equation}\label{app:solveGap}
We solve Equation~(\ref{eq:sigmaGap}), following
References~\cite{Son:1998uk,Park:1999bz}.
We split the integration range into, $0<p_0<k_0$ and $k_0<p_0<2\mu$.
Then the dominant contributions in the logarithms of the integrand are
$\ln\frac{2\mu p_0}{k_0^2}=2\ln\frac{2\mu}{k_0}-\ln\frac{2\mu}{p_0}$
for the former integration region and
$\ln\frac{2\mu}{p_0}$ for the latter.
We introduce the parameters
\begin{equation}
  x := \ln\frac{2\mu}{k_0}
  \;,\quad
  y := \ln\frac{2\mu}{p_0}
  \;,\quad
  x_0 := \ln\frac{2\mu}{\Sigma_0}
  \;,
\end{equation}
where $\Sigma_0 := \Sigma(0)$.
Then the gap equation is cast into the form
\begin{equation}
  \Sigma(x) = \frac{g^2}{8\pi^2}\frac{N_c^2-1}{N_c}
  \left(
        2x\int_x^{x_0}\Sigma(y)dy
        -\int_x^{x_0}y\Sigma(y)dy
        +\int_0^xy\Sigma(y)dy
    \right)
  \;.
\end{equation}
Note that we have $\Sigma(x_0)=-\Sigma(0)$.
We can take the derivative with respect to $x$ twice to get
\begin{equation}
  \Sigma'' = - \frac{g^2}{4\pi^2}\frac{N_c^2-1}{N_c}\Sigma
  \;,
\end{equation}
and the appropriate solution is
\begin{equation}
  \Sigma(x)=
  \Sigma_0\cos\left(x\sqrt{\frac{g^2}{4\pi^2}\frac{N_c^2-1}{N_c}}\right)
  \;,\quad\text{with}\quad
  x_0 = \pi\sqrt{\frac{4\pi^2}{g^2}\frac{N_c}{N_c^2-1}}
  \;.
\end{equation}
From the definition of $x_0$, we obtain
\begin{equation}
  \Sigma_0 = 
  2\mu\exp\left[-\pi\sqrt{\frac{4\pi^2}{g^2}\frac{N_c}{N_c^2-1}}\right]
  \;.
\end{equation}

\pagebreak


\end{document}